\documentclass[%
preprint,
 amsmath,amssymb,
 aps,
]{revtex4-2}
\usepackage[dvipsnames]{xcolor}
\definecolor{red}{RGB}{219, 1, 1}
\usepackage{graphicx}
\usepackage{dcolumn}
\usepackage{bm}

\begin{document}


\title{Evidence for large Rashba splitting in PtPb$_{4}$ from Angle-Resolved Photoemission Spectroscopy} 

\author{Kyungchan Lee$^{1,2}$, Daixiang Mou$^{1,2}$, Na Hyun Jo$^{1,2}$, Yun Wu$^{1,2}$, Benjamin Schrunk$^{1,2}$, John.M. Wilde$^{1,2}$, Andreas Kreyssig$^{1,2}$,  Amelia Estry$^{1}$, Sergey L. Bud'ko$^{1,2}$, Manh Cuong Nguyen$^{1,2}$, Lin-Lin Wang$^{1,2}$ Cai-Zhuang Wang$^{1,2}$, Kai-Ming Ho$^{1,2}$}

\author{Paul. C. Canfield$^{1,2}$}
 \altaffiliation[canfield@ameslab.gov]{}
\author{Adam Kaminski$^{1,2}$}%
 \email{adamkam@ameslab.gov}
\affiliation{%
 $^{1}$Ames Laboratory US Department of Energy, Ames, Iowa 50011, USA \\ 
 $^{2}$Department of Physics and Astronomy, Iowa State University, Ames, Iowa 50011, USA}%

\date{\today}

\begin{abstract}
 We studied the electronic structure of PtPb$_{4}$  using laser angle-resolved photoemission spectroscopy(ARPES) and density functional theory(DFT) calculations. This material is closely related to PtSn$_{4}$, which exhibits exotic topological properties such as Dirac node arcs. Fermi surface(FS) of PtPb$_{4}$ consists of two electron pockets at the center of the Brillouin zone(BZ) and several hole pockets around the zone boundaries. Our ARPES data reveals significant Rashba splitting at the $\Gamma$ point in agreement with DFT calculations. The presence of Rashba splitting may render this material of potential interest for spintronic applications.
 
\end{abstract}

\maketitle

\section{INTRODUCTION}

Recently, the search for new topological materials has resulted in a number of discoveries of very interesting materials, and has become a popular trend in condensed matter physics. Unique topological states such as spin-momentum locking\cite{kane2011TI_S_P_locking}, conductive surface state\cite{hasan2010colloquium}, back scattering suppression\cite{qi2010quantumbackscatt}, Fermi arc\cite{xu2015discovery_fermi_arc} and surface states protected by time-reversal symmetry\cite{hasan2010colloquium,moore2010birth} offer promise of a wide range of applications and significant advancement of computing technologies \cite{li2019dirac,belyakov2014heterogeneous,kuhmann1998pt,kempf1998thermodynamic,biggs2005hardening,wang2017room_TI_spintronics}. The ability to generate spin currents in topological materials is very important  for spintronic applications\cite{wang2017room_TI_spintronics}. The search for novel topological materials often starts with theoretical investigations such as DFT calculations later  confirmed by experimental studies\cite{ghosh2019saddle, xu2015discovery_fermi_arc, xu2015experimental_TaP, Pt2HgSe3_ARPES_cucchi2020bulk, theory_WSM}.

Recent measurements of the band structure in PtSn$_{4}$ revealed  Dirac node arcs which are with Dirac-like dispersions extending along one dimensional line instead of  having cylindrical symmetry\cite{wu2016}. PtSn$_{4}$ is composed of Sn-Pt-Sn layers along the $b$ axis with orthorhombic structure\cite{mun2012}. This is a rare example of the discovery of a topological material done by experiment rather than theory.

Building on our success with PtSn$_{4}$, we undertook an investigation of related materials to understand the origin of the Dirac node arc. We examined transport properties and band structures of PdSn$_{4}$, which belongs to the same structural family as PtSn$_{4}$\cite{jo2017}. In the case of PdSn$_{4}$, the Dirac node arc disappears but the single Dirac cone still persists upon replacing Pt with Pd.  This might indicate that the Dirac node arc could be sensitive to spin-orbit coupling(SOC).

To better understand the properties of PtSn$_{4}$ and structurally related ultra-heavy compounds with large SOC, we studied the electronic structure of  PtPb$_{4}$. Interestingly, this material is a superconductor with $T_C$ of 2.4~K \cite{gendron1962superconductivity}. In addition, it is an efficient electrocatalyst used in hydrogen fuel cells, which are important for renewable energy\cite{siahrostami2013enabling,li2019dirac}. PtPb$_{4}$ crystallizes in a tetragonal structure that is qualitatively very similar to the structure of PtSn$_{4}$. PtPb$_{4}$ is composed of Pb-Pt-Pb slabs that are simliar to the Sn-Pt-Sn slabs in PtSn$_{4}$. PtPb$_{4}$ has lattice constants $a$ $=$ $b$ $=$ 6.667~$\text{\AA}$ and $c$ $=$ 5.978~$\text{\AA}$ \cite{Villars2016:sm_isp_sd_1250530} and is predicted to be a strong topological insulator.  Although PtSn$_{4}$ and structurally related systems have been studied intensively, PtPb$_{4}$ has eluded  experimental studies of its electronic properties. Here, we report the electronic structure of PtPb$_{4}$ by using ultrahigh-resolution ARPES and DFT calculations. We found more than two electron pockets at the center of the BZ and several hole pockets around zone boundaries of each direction. Intriguingly, our ARPES data shows 2-fold symmetry even though the crystal structure is tetragonal. X-ray data and DFT calculations show that this is due to a modified stacking of the Pb-Pt-Pb layers of the  material.

\begin{figure*}[!htb]
\includegraphics[scale=0.8]{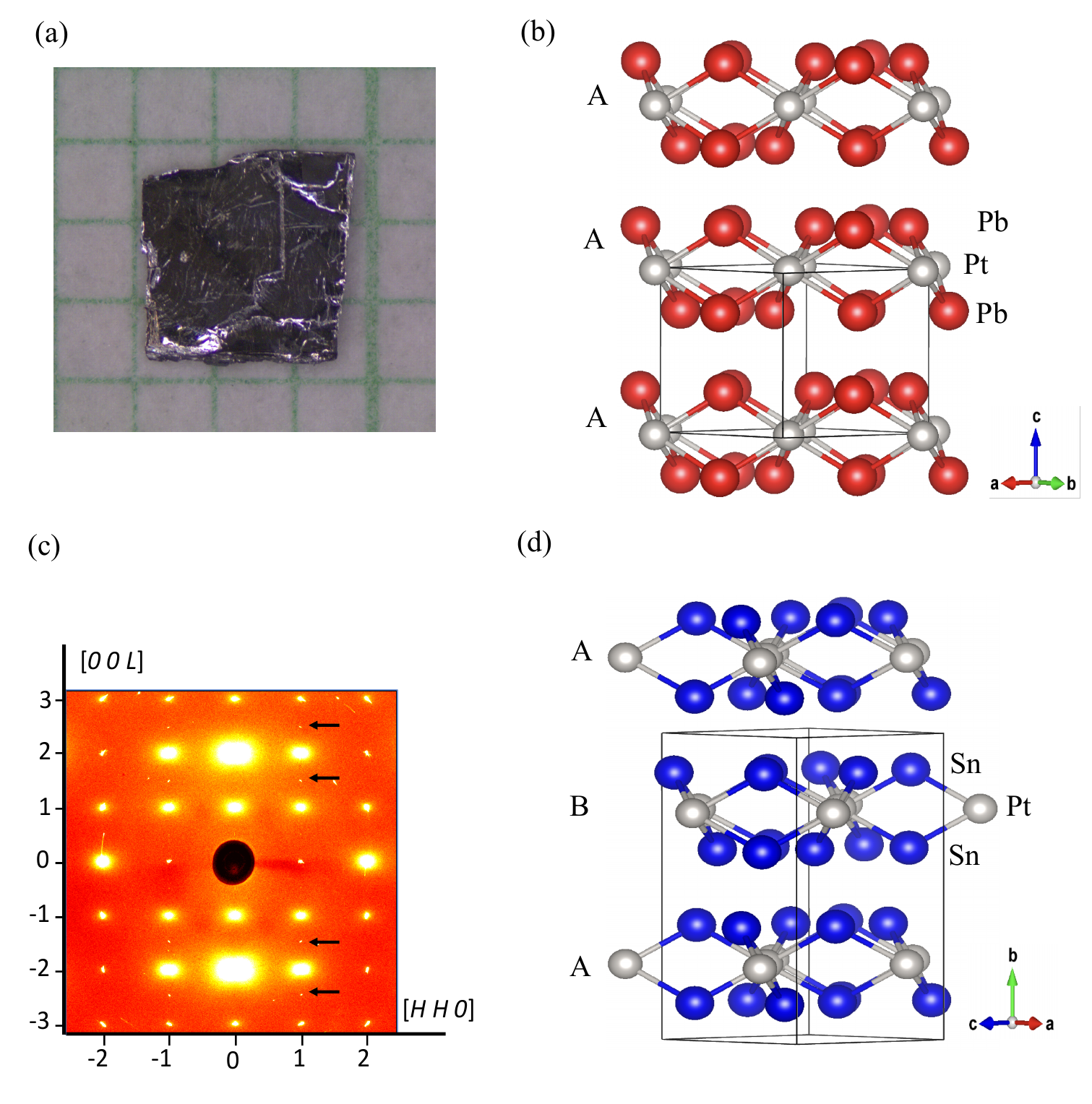}
\caption{\textbf{Single crystal of PtPb$_{4}$, X-ray scattering data of PtPb$_{4}$, crystal structure of PtPb$_{4}$ and PtSn$_{4}$} (\text{a}) photograph of PtPb$_{4}$ single crystal on a millimeter grid, (\text{b}) tP10 structure of PtPb$_{4}$ with the space group $P4/n b m$,  (\text{c}) high-Energy X-ray diffraction pattern of the $(H H L)$ plane, (\text{d}) oS20 structure of PtSn$_{4}$ with the space group $Ccce$.}
\end{figure*}

\begin{figure*}[!htb]
\includegraphics[scale=0.8]{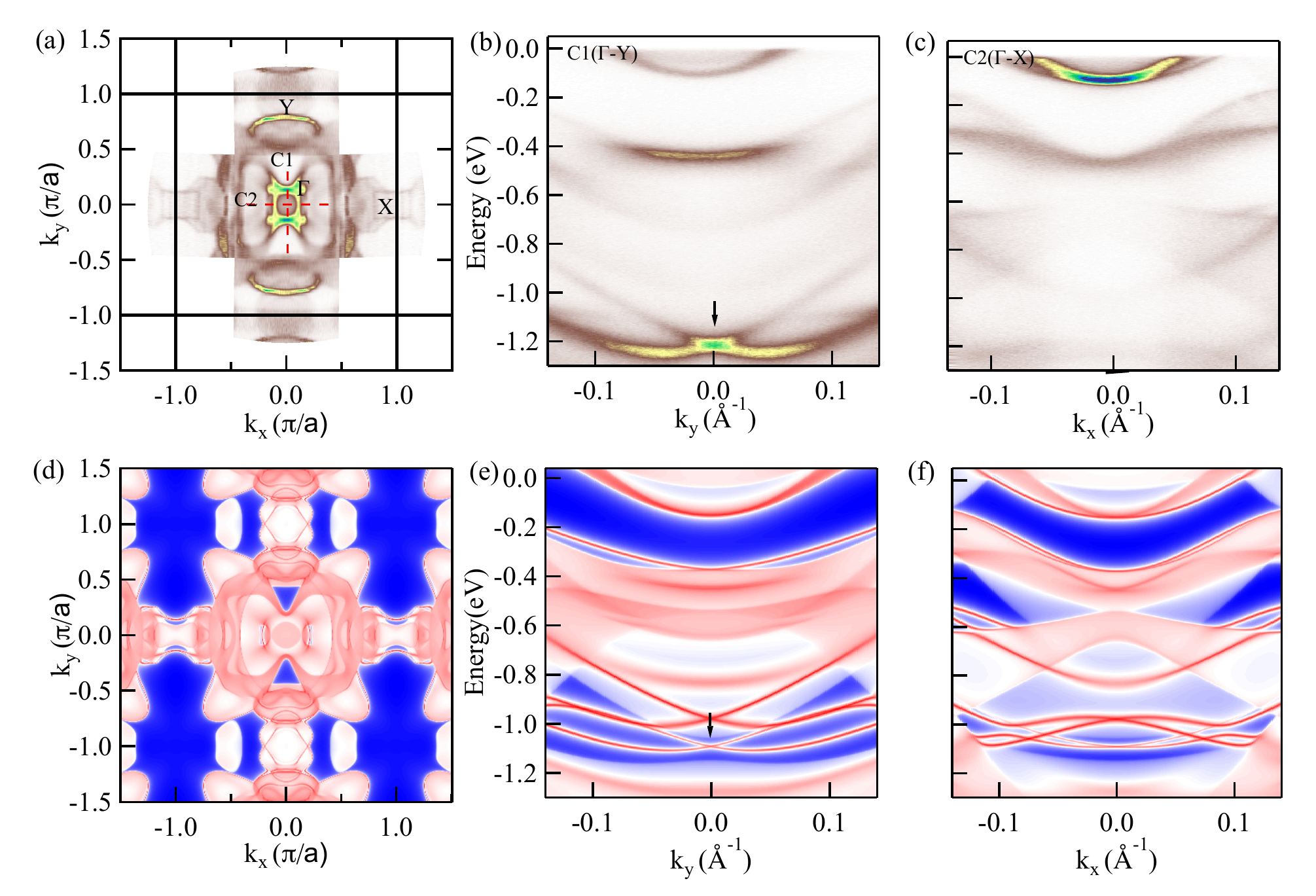}%
\caption{\textbf{FS, band dispersion of PtPb$_{4}$ at 40~K, calculated FS and band dispersion based on the oS20 PtSn$_{4}$ type structure} (\text{a}) FS plot of the ARPES intensity integrated within 10 meV of the chemical potential along $\Gamma$-X and $\Gamma$-Y. In this image plot, the brighter color represents higher intensity and the dark areas represent lower intensities. (\text{b}) band dispersion along $\Gamma$-Y direction, the center of RS is marked by black arrow.(\text{c}) band dispersion along $\Gamma$-X  (\text{d}) calculated FS,  (\text{e}) calculated band dispersion  along $\Gamma$-Y, (\text{i}) calculated band dispersion along $\Gamma$-X.}.
\label{fig:3}
\end{figure*}

Our measurements reveal large Rashba splitting (RS) at a binding energy of 1.3~eV. RS is a breaking of spin degeneracy as a result of SOC and inversion symmetry breaking. Although SOC appears various systems, it is stronger in materials including high-Z elements due to relatively larger spin-orbit parameter. Since RS is a spin-polarized surface state, it is expected to be useful in a generation of spin currents and manipulation of spin polarization by electric field for spintronic devices\cite{manchon2015new}. Despite decades of research, the operation of spintronic devices is very limited to low temperatures. Previous research  demonstrates that spin-injected field effect transistor (spin FET) controlled by gate voltage operates at 1.8~K\cite{koo2009control_lowtemp1} and the spin Hall effect transistor works at 4~K \cite{wunderlich2010spin_lowtemp2}. One of the major turning points in spintronic devices is using spin-polarized state of large RS. It was initially discussed in a wurtzite structure\cite{RS_ORGI_casella1960toroidal} and intensively studied in various  metals\cite{W_RS_rotenberg1999spin,Au_RS_lashell1996spin,Bi_RS_koroteev2004strong}, some alloys\cite{Bi_Ag_RS_ast2007giant,RS_alloy2}, thin films\cite{thin_film1_frantzeskakis2008tunable,thin_film2_he2008spin,yao2017direct_RS}, and semiconductors\cite{hatta2009_RS_SC, maass2016spin_RS_SC,ishizaka2011giant} which show giant Rashba splitting. 
Semiconductors typically show small RS, which is not suitable for room temperature spintronic devices. On the other hand, metals show relatively large RS but the spin polarization cannot be controlled by an external electric field. Our ARPES data and DFT calculations show that the RS in PtPb$_{4}$ is comparable to those of metal surface states. New generation spintronics devices working at room temperature need large splitting between spin polarized bands and demand tunability of chemical potential. Consequently, the spin polarized surface state in PtPb$_{4}$ gives us a chance to control the spin without applying magnetic fields at room temperature, paving a way for developing more efficient spintronics devices.

\section{METHODS}

Single crystals of PtPb$_{4}$ were grown out of Pb-rich binary melts.\,\cite{ASM2000} We put three different initial stoichiometries of Pt$_{13}$Pb$_{87}$, Pt$_{11}$Pb$_{89}$, and Pt$_{9}$Pb$_{91}$ into fritted alumina crucibles [CCS],\,\cite{Canfield2016} and then sealed it into amorphous silica tubes under partial Ar atmosphere. The ampoules were heated up to 600\,$^{\circ}$C, held there for 5 hours, rapidly cooled to 375\,$^{\circ}$C and then slowly cooled down to 310\,$^{\circ}$C over more than 100 hours, and then finally decanted using a centrifuge. 
The single crystalline samples have a clear plate like shape with a mirrored surface. Fig.1a shows a photograph of a single crystal of PtPb$_{4}$ on a millimeter scale. typical crystals have dimensions of 3 mm $\times$ 3 mm $\times$  0.5 mm. The crystallographic $c$ axis is perpendicular to the platelike plane. 

High-energy X-ray diffraction measurements were performed at station 6-ID-D at the Advanced Photon Source, Argonne National Laboratory. Measurements were made using 100 keV X-rays, with the incident beam direction normal to the $(HHL)$ reciprocal-lattice planes. Diffraction patterns were recorded using a MAR345 area detector. Unlike laboratory sources, high-energy X-rays ensure that the bulk of the sample is probed. By rocking the sample through small angular ranges about the axes perpendicular to the incident beam, we obtain an image of the reciprocal-lattice planes normal to the incident beam direction \cite{kreyssig2007crystallographic}. 

The density functional theory (DFT)\,\cite{kohn1965} calculations are performed using the Vienna Ab-initio Simulation Package (VASP) \cite{kresse1996} with projector-augmented wave (PAW) pseudo-potential method \cite{blochl1994}\cite{kresse1999} within generalized-gradient approximation (GGA) \cite{perdew1996}. The energy cutoff is 300\,eV and the Monkhost-Pack scheme \cite{monkhorst1976} is used for Brillouin zone sampling with a high quality \textit{\textbf{k}}-point mesh of 8\,$\times$\,8\,$\times$\,8 for tetragonal PtPb$_{4}$ and 8\,$\times$\,4\,$\times$\,8 for orthorhombic PtPb$_{4}$ structures, respectively. All structures are fully relaxed until the forces acting on each atom are smaller than 0.01\,eV/$\text{\AA}$ and all stress tensor elements are smaller than 1\,kbar (0.1\,GPa). Since both Pt and Pb are heavy elements, the relativistic effect is not negligible and is taken into account via spin-orbit coupling (SOC) calculations. The van der Waals (vdW) interaction is also taken into account by the DFT-D3 method \cite{grimme2010},\cite{grimme2011}. The surface formation energy is calculated by a slab model. The slab model consists 6 PtPb$_{4}$ layers with two identical surfaces [(001) for the tetragonal and (010) for the orthorhombic structures] and a vacuum region of more than 20\,$\text{\AA}$ in the direction perpendicular to the surfaces to avoid interaction between the slab  and its images due to periodic boundary conditions. A supercell of 2\,$\times$\,2 in the lateral directions is used containing 60 atoms. The 2 center layers are kept fixed at bulk atomic positions while atomic positions of the 2 outer most layers for each surface are relaxed. SOC and vdW interactions are included in the surface formation energy calculation.

ARPES measurements were carried out using a laboratory based tunable VUV laser. The ARPES system consists of a Scienta R8000 electron analyzer, picosecond Ti:Sapphire oscillator and fourth-harmonic generator \cite{jiang2014tunable}. All data were collected with 6.7~eV photon energy. Angular resolution was set at ∼0.1$^{\circ}$ and 1$^{\circ}$, respectively, and the energy resolution was set at 2 meV. The size of the photon beam on the sample was $\sim 30\,\mu$m. Samples were cleaved \textit{in-situ} at a base pressure lower than 1$\times$ 10$^{-10}$ Torr, 40 K and kept at the cleaving temperature throughout the measurement.

\begin{figure*}[!htb]
\includegraphics[scale=0.7]{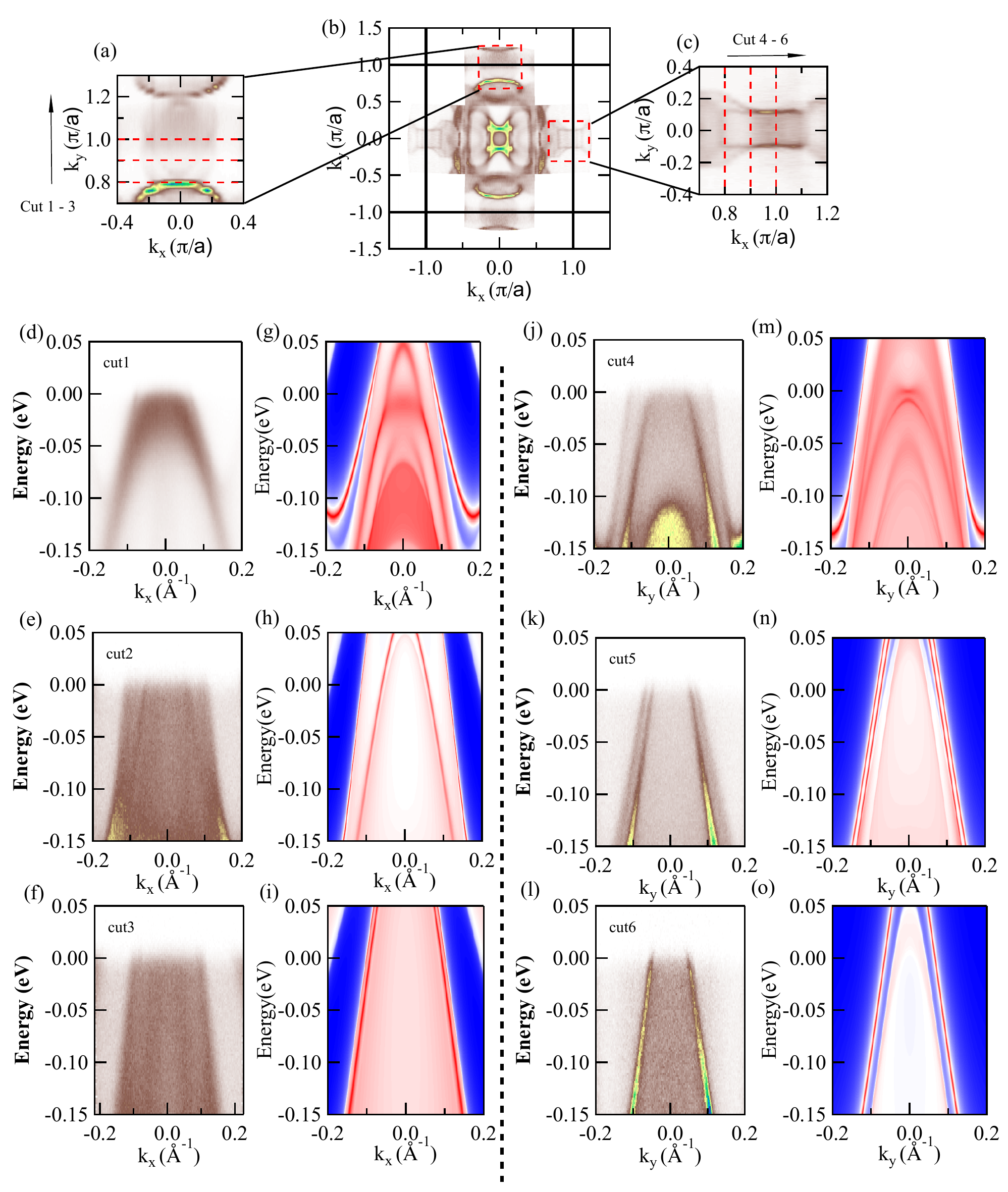}
\caption{\textbf{FS and band dispersion around the X and Y points at 40K. } (\text{a}) and (\text{c}) are magnified FS around the X and Y points. (\text{b}) FS surface of PtPb$_{4}$ at 40K, (\text{d})-(\text{f}) band dispersion around the Y point, (\text{g})-(\text{i}) DFT calculations corresponding energy dispersion.(\text{j})-(\text{l}) band dispersion around the X point and  (\text{m})-(\text{o})calculated band dispersion corresponding to (\text{j})-(\text{l}), respectively. }
\end{figure*}

\section{RESULTS AND DISCUSSION}

The previously reported crystal structure of PtPb$_{4}$ consists of stacked Pb-Pt-Pb layers along the $c$ axis as shown in Fig.1(b). The grey spheres represent Pt atoms, and red spheres are Pb atoms. High-energy X-ray diffraction data were taken on a single crystal sample of PtPb$_{4}$ as seen in Fig.1.(\text{c}) to search for any structural anomalies such as orthohombicity.  No peak broadening or splitting was observed, but additional peaks are clearly observed at half $L$ positions of $(1, 1, L)$ in Fig.1.(\text{c}), which are marked by black arrows. These peaks are forbidden for the reported space group of PtPb$_{4}$ and the resulting reflection conditions are not consistent with any tetragonal space group. However, several orthorhombic crystal structures like e.g. PtSn$_{4}$, would be consistent with the observed doubling of the unit cell in $c$ direction indicated by the additional half-integer Bragg peaks and the observed reflection conditions. Fig.1 (\text{d}) shows the most closely related  oS20 of PtSn$_{4}$ type structure by simply changing the stacking order. Grey balls represent Pt atoms and blue ones are Sn atoms.  

Our key experimental results are presented in Figs.2 and 3. We see a clear RS and find a surprising $C_{2}$ symmetry rather than $C_{4}$ symmetry in the $ab$ plane. Given the fact that we observe additional peaks at half $L$ positions of $(1, 1, L)$ in the X-ray data, we performed the ARPES measurement to understand the origin of these additional peaks. The FS and band dispersion along key directions in the BZ for PtPb$_{4}$ are shown in Figs.2(a)-(c).  Fig.2(a) shows the ARPES intensity integrated within 10 meV about the chemical potential, which roughly visualizes the FS. The FS consists of at least one electron pocket at the center of the BZ and is surrounded by several other electron pockets.  Previous results\cite{nordmark2002polymorphism} show that the crystal structure of PtPb$_{4}$ has AA stacking of PtPb$_{4}$ slabs along the tetragonal \textit{c}-axis with the space group $P4/n b m$. However, our X-ray and ARPES data point to the possibility of slight distortion of the crystal structure due to observed $C_{2}$ symmetry of the Fermi surface. Figs.2.(d)-(f) show the calculated FS and band dispersion based on oS20 structure with AB stacking of PtPb$_{4}$ slabs (with A and B being shifted in the basal plane with respect to each other) with the lattice constant doubled in $c$ direction. In the calculation, we used the PtSn$_{4}$ structure wtih 2 fold symmetry because it is shown to be energetically close to the PtPb$_{4}$ structure. Results seem to be in a good agreement with the ARPES data. The measured FS shows significant differences between the x and y directions and pockets around the X and Y points clearly have different shapes. Despite the tetragonal crystal structure of PtPb$_{4}$, the ARPES and X-ray scattering data indicate that the band structure at the surface may have different electronic properties than in the bulk. In order to understand this discrepancy between expectations based on the bulk crystal structure and ARPES measurements, we performed extensive band structure calculations.

\begin{table}[htbp]
	\centering
	\caption{Relative formation energy in meV/atom of tetragonal PtPb$_{4}$ to orthorhombic PtPb$_{4}$.}
\begin{tabular}{ccc}
          & GGA-PBE & GGA-PBEsol   \\
          \hline
w-SOC     & 2.2     & 1.3          \\
w-SOC-vdW & -1.3    & -4.3         \\
\end{tabular}
	\label{table}
\end{table}

Given that we see some evidence for 2-fold symmetry in our ARPES data, we used DFT calculations to determine how energetically removed  the possibility of tetragonal structure in PtPb$_{4}$. To be more specific, Table\,\ref{table} shows the energy difference of $E_{tet}$\,-\,$E_{ort}$ between tetragonal and orthorhombic PtPb$_{4}$ phases without the inclusion of the vdW interactions. It shows that the orthorhombic phase is more stable by -2.2\,meV/atom. We also perform a calculation using revised-GGA (PBEsol)\,\cite{perdew2008} to verify the calculated relative stability and find that the orthorhombic phase is also more stable within this approximation. We note that both tetragonal and orthorhombic phases are layered structures with a gap of $\sim$\,3.1\,~$\text{\AA}$ between Pb-terminated layers. Therefore, the vdW interactions would play an important role in determining the stability of different phases in this material. With the vdW interactions taken into account, it turns out that the tetragonal phase becomes more stable than  the orthorhombic phase by -1.3 and -4.3\,meV/atom for PBE and PBEsol GGA calculations. However, the energy differences, with and without vdW interaction correction, are very small within the order of a few K in term of thermal energy. In addition, the surface formation energies of these two phases are also very similar, which are 27.23 and 27.95~meV/${\AA^{2}}$ for orthorhombic and tetragonal phases, respectively. The orthorhombic phase has a slightly lower surface formation energy. Therefore, it is very likely that PtPb$_{4}$ is polymorphic and orthorhombic PtPb$_{4}$ can be stabilized close to the surface. Based on the X-ray data, DFT calculations and the ARPES measurement, the actual surface may  contain a mixture of domains with different terminations. 

In Fig.3, we focus on  the FS and band dispersion in the proximity of the X  and Y points.  An enlarged image from the red boxes in Fig.3(b) is shown in Fig.3(a) and Fig.3(c). Close to the X point, the FS has two parallel sections that consist of two merged hole-like bands as shown in Figs.3(c) and (j-l). Both bands are quite sharp and intense, which is consistent with a surface origin. The measured band  dispersion in this part of the zone agrees quite well with the DFT calculations shown in Figs.3(m)-(o). The parallel FS due to surface bands bare some resemblance to PtSn$_{4}$ and PdSn$_{4}$\cite{wu2016, jo2017}, although no topological features below $E_F$ are observed here; PtSn$_{4}$ has Dirac node arcs surface states but PtPb$_{4}$ shows a linear-like dispersion close to the X point.

The band structure near the Y point is shown in Figs.3 (b) and (d-f), while Figs.3 (g)-(i) show DFT calculations in the same part of the BZ. The ARPES intensity in this part of the BZ is weaker and bands are much broader, pointing to mostly a bulk origin that causes broadening due to the projection along $k_z$. The bands with strong surface components predicted by calculations Figs.3(h, i) are not well seen, which may indicate that matrix elements likely play a role in suppressing the signal.

We found a clear RS of the valence band. It depicts spin degeneracy lifting at surfaces or interfaces, where inversion symmetry is broken with spin-orbit interaction, which may lead to an asymmetric charge distribution. Relation between the breaking of spin degeneracy and the inversion symmetry can be described by the relativistic effect of moving electrons in a two-dimensional momentum space. The interaction between electron spin and effective magnetic field, originating from a surface electrostatic field, leads to a Zeeman splitting.  As a consequence, the degeneracy is lifted and chiral spin texture in band dispersion and maxima of the valence band or minima of conduction band are shifted from the symmetry point. Despite the reasonable success of initially proposed effective model\cite{RS_ORGI_casella1960toroidal, bychkov1984properties_RS_origin}, this model shows several limitations as well. The most important issue is the magnitude of the splitting. A result based on the effective model shows that the magnitude of splitting is, at least, several orders of magnitude smaller than that of previously observed data. On top of that, the effective model cannot explain the fact that SOC is stronger in high-Z elements because the strength of electric fields from the surfaces is roughly in the same order of magnitude. Recently, there are several attempts to explain this phenomenon by using numerical and analytic methods\cite{park2011orbital, nagano2009first_numerical_RS_origin}. Those results point out that RS may be related to the combination of an orbital angular momentum state and electron momentum that introduces asymmetric charge distribution. Although spin-orbit interaction occurs very often in various materials either in bulk or surface states, it is more pronounced in systems including high-Z elements due to relatively large spin-orbit parameter.

In spite of the breaking of $C_{4}$ symmetry, which may cause anisotropic spin-orbit coupling, we observed the RS in the $\Gamma$-Y direction. In some materials, it is not easy to distinguish a Rashba and Dresselhaus splitting experimentally. In order to understand the origin of the band splitting, we performed DFT calculations.  Based on the calculation results, we verify that the band splitting occurs at the surface state. The black arrow in Fig.2 (b) shows the position of the RS in the ARPES data. It clearly shows the RS with a valence band minimum (VBM) of $E_{\text{VEM}}$~=~1.3~eV at the momentum offsets of $\pm k_{0}$ with $k_{0}$ = 0.04 $\AA^{-1}$ around the $\Gamma$. This result shows that the RS in this material is several times bigger than that of Au(111)\cite{RS_Au2_lashell1996spin}, which is the initially reported ARPES measurement of a RS. In our ARPES data, the VBM is slightly lower than predicted by calculation. Fig.2 (c) shows the band dispersion along the $\Gamma$-X direction. We find at least one electron pocket in this direction. Figs.2 (e) and (f) show DFT calculations of $\Gamma$-Y and $\Gamma$-X cuts and bare resemblance to the  ARPES data modulo a small shift in the energy, which is quite typical.

The outstanding issue is the difference between the observed two-fold symmetry of the band structure and the four fold symmetry of the crystal structure of PtPb$_{4}$. Our DFT calculations and X-ray data led us to conclude that the symmetry breaking may be caused by an  ABAB stacking sequence; the orthorhombic structure is more favorable. Lastly, we observed a RS, at the center of the BZ, which typically appears at the surface or interface of materials. Even though this effect occurs below $E_F$, it may be possible by either chemical substitution or gating to move the chemical potential down to utilize this effect in a generation of spin currents at room temperature.  This may have interesting consequences not only due to its fundamental importance but also its possibility of application for spintronic devices. 

\section{ACKNOWLEDGMENTS}

This work was supported by the U.S. Department of Energy, Office of Science, Basic Energy Sciences, Materials Science and Engineering Division. Ames Laboratory is operated for the U.S. Department of Energy by Iowa State University under contract No. DE-AC02-07CH11358. K. L. was supported by CEM, an NSF MRSEC, under grant DMR-1420451. This research used resources of the Advanced Photon Source, a U.S. Department of Energy (DOE) Office of Science User Facility operated for the DOE Office of Science by Argonne National Laboratory under Contract No. DE-AC02-06CH11357.

\appendix

\clearpage

\nocite{*}

\providecommand{\noopsort}[1]{}\providecommand{\singleletter}[1]{#1}%
%


\end{document}